%%%%%%%%
%
% Following is the Tex source code for 
%
%  Negative Energy Densities and the Limit of Classical Space-Time,
%  by Adam Helfer
%%%%%%%%
%
% ************************************************************
%
% The next two lines define the symbols
% \gtrsim and \lesssim  (greter than/less than of order of magnitude).
% 
% If these symbols are already available, simply replace these
% lines with \def\gtrsim{whatever your macroname is for this}
%            \def\lesssim{whatever your macroname is for this}
% ************************************************************
\def\gtrsim{{\buildrel >\over\sim}}

%%%%%%%%%%%%%%%%%%%%%%%%%%%%%%%%%%%%%%%%%%%%%%%%%%%%%%%%%%%%%%%%%%%%
\baselineskip 22pt
\def\vsk{\vskip 11pt}
\newcount\seccount
\seccount=0
\newcount\subseccount
\subseccount=0
\newcount\fnotecount
\fnotecount=0
\def\fnote#1{\global\advance\fnotecount by
1\footnote{${}^{\the\fnotecount}$}{#1}}
\newcount\EEK
\EEK=0
\def\eek{\global\advance\EEK by 1\eqno(\the\EEK )}
\def\sec#1{\global\subseccount=0\global\advance\seccount by
1\vsk\vsk\noindent\bf\the\seccount .  #1\rm\vsk}
\def\subsec#1{\global\advance\subseccount by
1\vsk\noindent\bf\the\seccount .\the\subseccount .  #1\rm\vsk}
\def\T{{\widehat T}}
\def\H{{\widehat H}}
\seccount=-1
\fnotecount=-1

\def\references{%
\vfill\eject
\vsk\vsk\centerline{\bf References}
\vsk
\frenchspacing
\parshape=0\global\parindent=0pt\relax
\everypar{\hangafter=1\hangindent=2pc\relax}%
}

{\centerline{{\bf Negative Energies and the Limit of Classical
Space--Time}\footnote{${}^*$}{This essay received an ``honorable 
mention'' from the Gravity Research Foundation, 1998 --  Ed.}}}
\vsk
\centerline{Adam D. Helfer}
\centerline{Department of Mathematics, University of Missouri, Columbia, 
Missouri 65211, U.S.A.}%

\vskip .5in
\centerline{\it Summary}
Relativistic quantum field theories predict negative energy densities,
contravening a basic tenet of classical physics and a fundamental
hypothesis of the deepest results in classical general relativity. 
These densities may be sources for exotic
general--relativistic effects, and may also lead to pathologies.

Combining Ford's ``quantum inequality'' with quantum restrictions on
measuring devices,
I present an argument that these densities nevertheless satisfy a sort
of ``operational'' positivity:  the energy in a region, plus the energy
of an isolated device designed to detect or trap the exotic energy, must
be non--negative.  This will suppress at least some pathological effects.

If we suppose also that Einstein's field equation holds, then no
local observer can measure the geometry of a negative energy density
regime accurately enough to infer a negative
energy density from the curvature.  This means that the physics of a
negative energy regime cannot be adequately modeled by a classical
space--time.

\vfill\eject

\sec{Preface}

What follows is the essay I wrote for the Gravity Research Foundation's
1998 prize competition.  The editor and referee have suggested that I
add a few remarks to amplify some issues and to place the work in
context.  I am grateful for the opportunity to do so, as constraints of
space made this impossible in the original paper.  Indeed, there is a
temptation to write a Shavian preface twice as long as the piece.
However, I can resist anything --- except temptation.

This paper has very little speculation in it.  The overall
approach is to take conventional quantum field theory in curved
space--time, and conventional quantum mechanics, and apply them at
conventional scales.  Thus I do not attempt to learn how physics might
be modified at, for example, the Planck scale, or what exotic effects
might be produced by string theory.  I do not suppose my present
approach will apply directly to such extreme regimes, which
have been investigated by other authors [0].  I do not use any version
of ``quantum gravity'' in the sense this term is usually understood. 
Still, I am led to infer a quantum character for space--time in certain
regimes --- by showing that a classical model is not tenable.

The regimes in question are the negative energy--density configurations
arising in relativistic quantum field theories.  These regimes are
predicted to occur \it generically.  \rm  Thus a central question is,
Why are negative energy density effects not pervasive?  Part of the
answer, I suggest, is a restriction on quantum measurement, deducible
from known physics but not previously considered.  This is the
``operational weak energy condition,'' which requires that the energy
in a regime, plus the energy of an isolated device  in that regime
constructed to measure or trap that energy, be non--negative.

It is worth emphasizing that this condition is a restriction on the
ability of a measuring device \it in a specific space--time region \rm
to achieve certain results.  This sort of restriction does not seem to
have received much attention (beyond the casuality requirements
embodied in the space--like commutation of field operators).  

I have avoided talking about ``interpretations'' of quantum
mechanics, as those who feel strongly about the subject will certainly
draw therir own conclusions of my work's significance.  However,
whatever view one has, I should like to reiterate the point of the
previous paragraph:  the present analysis can only be accomodated by
considering \it what \rm devices might measure the energy density, and
\it where \rm they are located.  It could not be accomodated within any
set of assumptions which \it presume \rm the existence of ideal
measuring devices (devices measuring arbitrary self--adjoint operators
without otherwise interfering with the system), or without considering
where the devices are situated.

The restriction on measurability comes from the restrictions on the
mass of a clock with given resolving time (which may be attributed to
Bohr, Einstein, Schr\"odinger, Salecker and Wigner); the operational
positivity of energy density comes from combining this with ``quantum
inequalities'' of Ford and Roman, which limit the times for which
negative energy densities may persist.  There has been previous work on
limitations of quantum measurements of position based on related
considerations [0].  On the other hand, the present work contrasts with
restrictions on measurements which have been proposed to arise from
deformations of the canonical commutation relations.

Assuming Einstein's field equation, the geometry is related to the
energy density, and so the restrictions uncovered forbid direct
measurements of the curvature of space--time in regimes where the
energy density is negative (at least, of those predicted by
conventional quantum field theory).\fnote{It is here that the link
between space--time geometry and quantum measurement theory is made.
In this sense I am making a ``quantum gravitational'' assumption.} What
this means is that one cannot verify by \it direct local \rm means that
the space--time geometry of a negative energy regime ``is really
there.'' (There could be indirect evidence for it, however.) The
geometry can only be measured locally by introducing devices so massive that
they swamp the negative--energy effects.

The details of these arguments are set out more fully in [7].
Roughly speaking, at the simplest level, the usual ``test particle''
thought--experiments to measure the geometry of space--time cannot
succeed in negative energy--density regimes, because particles with
Compton wavelengths short enough to accurately probe the
curvatures driven by negative energies turn out to be massive enough to
destroy that negativity.  

This undoubtedly has a queer sound to it, and it takes some work to
understand what it means.  In this essay I have given a simple example,
relevant to Cosmic Censorship, in which it is easy to see the
physical consequences of the restrictions.  However, a more detailed
analysis addresses the issue of to what extent we may say that there is
a classical space--time in a negative energy--density regime and to what
extent we are forced to impute a quantum character to the geometry. 
Such an analysis will be found in [7], where it is shown that it is
very hard to ascribe any meaningful classical geometry to these regimes.
Even if we abandon the test--particle picture, and attempt to take into
account the interaction of the measuring device and the field, there
are considerable obstacles to giving a meaningful classical character to
the negative energy--density regime.
The concern that quantum measurement processes might be incompatible
with general relativity has been raised earlier [0'].

\sec{Introduction}

All known classical forms of matter have positive energy
density.  Indeed, that the energy of a system be positive
(or, in the non--relativistic context, bounded below) controls
some of the most basic aspects of its behavior.  It is what
makes thermodynamic equilibrium and the laws of thermodynamics
possible; and it is required for dynamic stability.  In
General Relativity, positivity of energy density (and similar
``energy conditions'') are
at the heart of the deepest results in the field:  the singularity
theorems, the area theorem for black holes, and the
positivity--of--total--energy theorems.

Yet it is well--known that in \it some \rm senses quantum
fields may have negative energy densities.  This has motivated
the search for exotic effects driven by such densities.  Serious
workers have investigated possible thermodynamic paradoxes [1], as
well as ``traversable worm holes,'' ``warp drives'' and ``time machines''
[2].
Of course, it is understood that such work is very speculative.

Curiously, many of the initial investigations have been followed by
others suggesting that these exotic effects are difficult or impossible
to attain [3].  Sometimes one needs devices at something like the Planck
scale to create the negative densities demanded by particular
hypothesized applications.  In other cases there are problems in
usefully controlling the effects.  One is led to wonder if these
results are manifestations of some deeper principle suppressing exotic
negative energy effects.

A recent result suggests that such a 
principle exists.  It has been shown that, in generic space--times,
the energy density and even total energy operators are \it always \rm
unbounded below, and the set of states on which their expectations are
$-\infty$ is \it dense \rm in the Hilbert space [4].\fnote{Precisely,
if $\T _{ab}$ is the renormalized stress--energy, then the Hamiltonian
$\H (\xi )=\int \T _{ab}\, \xi ^a\, d\Sigma ^b$
associated to a timelike vector field $\xi ^a$
is unbounded below, and the states on which its expectation is $-\infty$
are dense,
unless $\xi ^a$ can be chosen to be a Killing field
at $\Sigma$.}
In such a 
setting, it is hard to see how exotic negative energy effects could be
avoided, unless there is a general principle which tends to suppress
them.  What, for example, prevents an ordinary particle from absorbing
negative energy and becoming a tachyon?  If there were \it any \rm
cross--section for such a process, it is hard to see how, given the
pervasiveness of very negative energy states, we would have failed
to see it.
\vfill\eject

\sec{An ``Operational'' Weak Energy Condition}

Important restrictions on negative energy densities were
discovered by Ford [5].  The \it quantum inequalities \rm 
limit the time a negative energy density may persist.
For example, for the Klein--Gordon field on Minkowski space, 
$$\langle\Phi|\int _{-\infty}^\infty \T _{00}(t,0,0,0)\, b(t)\, dt
  |\Phi\rangle\geq -(3/32\pi ^2)\hbar c/(ct_0)^4
  \qquad\hbox{for any normalized } |\Phi\rangle \, ,\eek$$
where the sampling function $b(t)=(t_0/\pi )/(t^2+t_0^2)$ has area unity
and characteristic width $\sim t_0$.  
While no similar results have been rigorously proved for generic curved
four--dimensional
space--times, there are good reasons for thinking that they will
hold.  I shall assume, as do most workers, that they do.

A crucial issue is pointed up by the quantum inequalities.  
When negative energy density effects are important, one cannot
simply speak of the ``energy in a region;'' one must include a notion of
the time scale over which the energy is averaged.  The same state could
have very negative energies when measured over a short time, and less
negative energies when averaged over a longer time.  We shall speak of
the energy content of a \it regime, \rm understanding that
this refers not just to a region of space, but also to a temporal
averaging.  

We now combine the quantum inequalities with quantum
restrictions on a device which might measure or trap energy.
Consider an \it isolated \rm device designed to measure or trap a negative
energy density.  Since the 
negative energy density can persist for only a limited time, the device
must have a clock which turns it on and off, say on a time $\sim t_0$.
A clock which resolves times of order $t_0$ must
have rest--energy $\gtrsim \hbar /t_0$~[6].  On the other hand, the total 
negative energy detected or absorbed is restricted by the quantum
inequalities and causality:
$$|E _{\rm neg}|\leq (4/3)\pi (ct_0)^3\cdot(3/
  32\pi ^2)\hbar c/(ct_0)^4 =(8\pi )^{-1}
  \hbar /t_0\, .$$
Thus the energy of the measuring device must be greater than the negative
energy detected.

This may be called an \it operational weak energy condition \rm for
the Klein--Gordon field in Minkowski space:  the
energy in a regime, plus the energy of an isolated device constructed to
measure or trap that energy, must be non--negative. 

I suggest that the operational weak energy condition is valid
generally, for all quantum fields.  I should emphasize that the result
has not been proved with mathematical rigor even for the Klein--Gordon
field in Minkowksi space.\fnote{The argument as given
depends on a special choice of sampling function, and the inequality
$E_{\rm clock}\gtrsim \hbar /t_0$ is only known to hold in an 
order--of--magnitude sense.}  
Nevertheless, the result is so suggestive, and the
factor $8\pi$ so in excess of unity, that it seems at least worth
considering.

A thought experiment to measure energy density in a region
gravitationally has been investigated in some detail, and the
operational weak energy condition appears explicitly [7].  Indeed, at
least for the device considered, timing errors prevent one from coming
close to saturating the condition unless the Planck scale is approached.

It is not clear that this operational weak energy condition will rule
out all pathological effects, let alone exotica.  Each must be
examined carefully.  Certainly the
condition would prevent an ordinary particle from absorbing negative
energy and becoming a tachyon. 

We shall see however that the consequences of the 
condition are of interest whether or not it resolves all the
pathologies.

\sec{Limitations of a Classical Model for Space--Time}

The operational weak energy condition would imply that negative
energy densities have, in some sense, a will--o'--the--wisp character.
While they might be definitely predicted by theory (as, for example,
between the plates of a Casimir apparatus), they cannot be confirmed
by a \it direct local \rm experimental meausurement --- for this
would always require a device 
massive enough to swamp the negative energy density.

To begin to understand the physical significance of this, let us consider
a situation relevant to the Cosmic Censorship Conjecture.  
Suppose a singularity is present in a negative energy density regime.
Could this be visible from infinity?

If the operational weak energy condition holds, then, while there might
be \it mathematical \rm null geodesics escaping from the singular region
to infinity, these geodesics could not carry \it physical \rm photons
of short enough wavelength to give a detailed image
of the singular region.  This is because any measurement of the 
singular region accurate enough to measure the curvature would imply
a measurement of the energy density, by Einstein's equation.  And if
the energy density is negative, this is forbidden by the operational 
weak energy condition.

While this argument has a little interest as circumstantial support for
the Cosmic Censorship Conjecture, it is much more important in that
it shows that there may be a clear distinction between the \it mathematical
\rm model of a classical space--time and the \it physical \rm
possibilities for measurement of geometry and transmission of information
by signals.  This is in fact a general feature of negative energy density
regimes.

We may say that a space--time region is modeled classically
(to a desired
accuracy) if it is possible in principle to introduce
test particles (i.e., particles not interfering with the measurement)
whose trajectories can be measured to determine the geometry of 
the region (to the desired accuracy).\fnote{The term ``particle'' does not
imply a point mass, but only an object with some degree of localizability.}

Classical existence in this sense is forbidden by the operational
weak energy condition in negative energy density regimes, whenever the
accuracy is enough to infer the energy density from Einstein's equation.
Whether one is bold enough to say that such a regime is
quantum space--time is a matter of temperament.  However, the inadequacy
of the classical model is fairly clear.  Picturesquely:  

\itemitem{} \it A space--time regime can have a negative energy density only 
if no one is there to measure it! \rm

\sec{Conclusion}

General Relativity is ineffably beautiful, and 
only with the greatest caution should we seek to move beyond it.
Yet the principle on which Einstein founded this theory, and which was
essential to the development of quantum theory, was operationalism:
that a theory should be formulated in terms of physical observables.
Einstein himself was sensible that the geometry of space--time in the
quantum regime would ultimately have to be justified operationally:

\item{} It is true that this proposed physical interpretation of
geometry breaks down when applied
immediately to spaces of sub--molecular order of magnitude....
Success alone can decide as to the justification of [attempts to do so]....
It might possibly turn out that this extrapolation has no better
warrant than the extrapolation of the idea of temperature to parts
of a body of molecular order of magnitude. [8]

\references

[0] See G. Amelino--Camelia, \it Mod. Phys. Lett. \rm\bf A9 \rm (1994) 335,
\bf A9 \rm (1994) 3415, \bf A12 \rm (1997) 1387, and references therein.

[0'] D. V. Ahluwalia, \it Phys. Lett. \rm\bf B339 \rm (1994) 301.

[1] P. C. W. Davies, \it Phys. Lett. \rm\bf 113B \rm (1982) 215.

[2] M. Morris and K. Thorne \it Am. J. Phys. \rm\bf 56 \rm (1988) 395;
M. Morris, K. Thorne and Y. Yurtsever \it Phys. Rev. Lett. \rm\bf 61 \rm
(1988) 1446;
M. Alcubierre, \it Class. Quant. Grav. \rm\bf 11 \rm (1994) L73.
 
[3] P. G. Grove, \it Class. Quant. Grav. \rm\bf 5 \rm (1988) 1381;
L. H. Ford, \it Phys. Rev. \rm\bf D43\rm (1991) 3972;
L. H. Ford and T. A. Roman, \it Phys. Rev. \rm\bf D53 \rm (1996) 5496;
M. J. Pfenning and L. H. Ford, The unphysical nature of
`warp drive,' preprint gr--qc/9702026.

[4] A. D. Helfer, \it Class. Quant. Grav. \rm\bf 13 \rm (1996) L129.

[5] L. H. Ford and T. A. Roman, \it Phys. Rev. \rm\bf D55 \rm (1997)
2082.

[6] H. Salecker and E. Wigner, \it Phys. Rev. \rm\bf 109 \rm (1958) 571.

[7] A. D. Helfer, `Operational' Energy Conditions, \it Class. Quant.
Grav., \rm \bf 15 \rm (1998) 1169, gr--qc/9709047.

[8] A. Einstein, Geometry and Experience, address to the Prussian
Academy of Sciences, January 27, 1921,
in \it Sidelights on Relativity, \rm 
Dover.
%http://ourworld.compuserve.com/homepages/eric_baird/ae_geoex.htm
\bye